\documentclass[aps,pre,twocolumn,float]{revtex4}
\usepackage{amsmath,bm,epsfig}

\def\Fbox#1{\vskip1ex\hbox to 8.5cm{\hfil\fboxsep0.3cm\fbox{%
  \parbox{8.0cm}{#1}}\hfil}\vskip1ex\noindent}  

\newcommand{\B}[1]{{\bm{#1}}}
\newcommand{\C}[1]{{\mathcal{#1}}}    
\let \= \equiv \let\*\cdot \let\~\widetilde \let\^\widehat \let\-\overline

\begin{document}
\title{The Effective Temperature in Elasto-Plasticity of Amorphous Solids}
\author{Laurent Bou\'e, H.G.E. Hentschel$^*$, Itamar Procaccia, Ido Regev and Jacques Zylberg}
\affiliation{Department of Chemical Physics, The Weizmann
Institute of Science, Rehovot 76100, Israel. \\
$^*$ Dept of Physics, Emory University, Atlanta Ga. 30322. }
\date{\today}
\begin{abstract}
An effective temperature $T_{\rm eff}$ which differs from the bath temperature is believed to play an essential role in the theory of elasto-plasticity of amorphous solids. The definition of a measurable $T_{\rm eff}$ in the literature on
sheared solids suffers however from being connected to a fluctuation-dissipation theorem which is correct only in equilibrium. Here we introduce a natural definition of $T_{\rm eff}$ based on measurable structural features without recourse to any questionable assumption. The value of $T_{\rm eff}$ is connected, using theory and scaling concepts, to the flow stress and the mean energy that characterize the elasto-plastic flow.
\end{abstract}
\maketitle
{\bf Introduction}: Amorphous solids form when super-cooled liquids are further cooled below the glass transition.
While indistinguishable in their microscopic disorder from fluids, amorphous solids exhibit, in contradistinction from
fluids, a yields stress below which they respond elastically to external strains; fluids flow under any external strain.
Once the amorphous solid is subject to large enough strains such that the response of the internal stress exceeds the
yield stress, it can flow plastically in a manner that depends on the temperature, the shear rate, the density etc. While there have been many attempts to present phenomenological equations to describe the rheology and the constitutive relations of such elasto-plastic flows  \cite{79AK,79Arg,82AS,98FL,98Sol,07BLP,09BCA}, to this date none of these attempts has gained universal acceptance. In fact, there is no complete agreement even on the field variables, or `order parameters' which are necessary to close a complete set of equations.

Among the more interesting ideas for order parameters stands the proposition that such elasto-plastic flows exhibit
two different temperatures, the regular temperature $T$ that relates to the mean velocity of the particles forming
the amorphous solid (and, naturally, to the heat bath to which the system is coupled), and an `effective temperature' $T_{\rm eff}$ that has to do with some `noise' \cite{97SLHC} or `compositional properties' of the material \cite{09BL}.
The concept of effective temperature was originally formulated in Ref. \cite{89Ed} in order to describe the macroscopic properties of granular matter.  Although these systems are athermal in the sense that they cannot evolve under normal temperature conditions, they may be driven by an external force which establishes an equilibrium like behavior described by an effective temperature related to the intensity of the ``tapping''.  These ideas have since been  applied in other kinds of mechanically driven systems such as disordered elastic structures \cite{98Cug,07BK}. For sheared amorphous solids the definition of this effective temperature was never entirely convincing, being connected either to the density of ill-defined `shear-transformation zones' \cite{07BLP,09BL} or related to some `fluctuation-dissipation theorem' \cite{02BB,04OLN,07HL} that should be exact only in equilibrium \cite{Chetrite}. The aim of this Letter is to announce a simulational discovery of very sharp and direct meaning to an effective temperature in a number of simple computer models of elasto-plasticity. This effective temperature has an obvious connection to the compositional disorder in the material. Moreover, it naturally identifies with the regular equilibrium temperature in the super cooled liquids. In this way it allows a smooth description that unites the super-cooled regime with the amorphous-solid regime, something that is certainly lacking in many phenomenological descriptions. To define $T_{\rm eff}$ we need first to recall some recent advances in describing the super-cooled equilibrium regime.

{\bf Up-scaling in the super-cooled regime:} In a series of recent papers (cf. \cite{series} and in particular
 \cite{09BLPZ}) it was proposed that the
scenario of the glass transition, including the astonishingly rapid slowing down of the dynamics in a short range of temperatures, is usefully encoded by the temperature dependence of the concentrations of a finite set of quasi-species which can be indexed by $1,2,\cdots, n$. The precise nature of these quasi-species may change from model to model, but they are always formed by particles and their nearest neighbors. The main advantage of
these quasi-species is that they obey a discrete statistical mechanics, in the sense that
their temperature-dependent concentrations $\langle C_i\rangle(T)$, are determined by a set of degeneracies $g_i$ and enthalpies $\C H_i$ such that
\begin{equation}
\langle C_i\rangle(T) = \frac{g_i e^{-\C H_i/k_B T }}{\sum_{i=1}^n g_i e^{-\C H_i/k_B T}} \ . \label{CiT}
\end{equation}

Obviously, if such a simple description is available, we can predict which quasi-species will
be there when the temperature is high, and which will remain when the temperature decreases: only
those with lowest free energy remain at low temperatures. Indeed, simulations show how the concentrations of some quasi-species decrease, some increase, and some start increasing and then decrease when temperature is lowered, according to their degeneracy and enthalpy. Fluidity (or short relaxation times) is therefore correlated with high concentrations of quasi-species whose free energy is high, and solidity (or long relaxation times) is correlated with high concentrations of quasi-species whose free energy is low. This qualitative observation was made quantitative by
noting the concentrations of those quasi-species that tend to disappear
when the temperature is lowered, and summing these concentration to what was called the `liquid-like' concentration $\langle C_\ell\rangle (T)$. The inverse of this concentration provides a
length-scale (the typical distance between `fluid' quasi-species):
\begin{equation}
\xi(T)\equiv [\langle C_\ell\rangle (T)]^{-1/d}\ , \quad \xi\to \infty ~{\rm when}~ T\to 0, \label{xi}
\end{equation}
where $d$ is the space dimension. It was amply demonstrated on a large variety of models that the
relaxation time $\tau_\alpha(T)$ measured using correlation functions in the super-cooled regime
is determined by this diverging scale according to
\begin{equation}
\tau_\alpha = \tau_0 e^{\mu \xi(T)/T} \ , \label{tauxi}
\end{equation}
where $\mu$ is a typical free energy per particle and $\tau_0$ a microscopic (cage) time. In contradistinction with the Vogel-Fulcher and Adam-Gibbs fits, Eqs. (\ref{xi}) and (\ref{tauxi}) imply that there is no singularity associated with the glass transition at any temperature other than $T=0$, as was explained in \cite{07EP}. The discovery that we announce here is that the crucial statistical mechanical relation
Eq. (\ref{CiT}) can remain correct and very useful, with $T$ replaced by $T_{\rm eff}$, also
in the elasto-plastic regime of the amorphous solids that form at ultra-low bath temperatures. In other words, $T_{\rm eff}$ exists and it determines the compositional disorder of the material.

{\bf Two Models}: We present the new findings using two different models of glass formation in two-dimensions, the first being the Shintani-Tanaka model \cite{06ST}, and the second the so-called `hump model' which was inspired by \cite{89Dzu} and analyzed in Ref. \cite{09BLPZ}. The Shintani-Tanaka model has $N$ identical particles of mass
$m$; each of the particles carries a unit vector $\B u_i$ that can rotate on the unit circle. The particles interact via the potential $U(r_{ij} , \theta_i, \theta_j) = \bar U(r_{ij} ) + \Delta U(r_{ij} , \theta_i, \theta_j)$. Here $\bar U(r_{ij} )$ is the standard isotropic Lennard-Jones 12-6 potential, whereas the anisotropic part $\Delta U(r_{ij} , \theta_i, \theta_j)$ is chosen such as to favor local organization of the unit vectors in a five-fold symmetry, to frustrate crystallization. For full details of this model the reader is referred to Refs. \cite{06ST,07ILLP,09LPR}; here it suffices to know that with the parameters chosen in Ref. \cite{06ST} the model crystallizes upon cooling
for $\Delta<0.6$ whereas for larger values of $\Delta$ the model exhibits all the standard features of the glass transition, including a spectacular slowing down of the decay of the correlation functions of the unit vectors
$C_R (t) \equiv (1/N) \sum_i \langle \B u_i(t)\cdot \B u_i(0)\rangle$ which is very well described by Eq. (\ref{tauxi}).

\begin{figure}
\begin{center}
\includegraphics[width=0.6\linewidth]{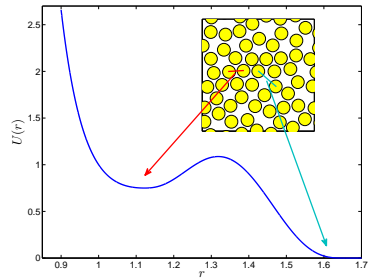}
\caption{The pair-wise potential for the hump model. In the inset we show a snapshot of the position
of the point-particles (the circles represent the finite range of interaction, cf. \cite{09BLPZ}).}
\label{fighump}
\end{center}
\end{figure}

The `hump model' again employs $N$ identical particles interacting via a potential that is constructed as a piecewise function consisting of the repulsive part of a standard 12-6 Lennard--Jones
potential connected at $r_0 = 2^{1/6}\sigma$ to a polynomial
interaction $P(x) = \sum_i a_i x^i$.  The $a_i$'s are tuned \cite{09BLPZ} so that
$P(x)$ displays a peak at $r = r_{\mbox{\tiny hump}}$ and also such
that there is a smooth continuity (up to second derivatives) with
the Lennard--Jones interaction at $U(r_0) = \epsilon h_0$ as well
as with the cut-off interaction range $U(r_{\star}) = 0$.  The
interaction potential for the hump model is shown in Fig. \ref{fighump}
Note that the two typical distances that are defined by this potential, i.e. the distance at the minimum $r_{\rm min}$ and the cutoff scale $r_\star$, appear
explicitly in the amorphous arrangement of the particles in the supercooled liquid, as shown in the inset in Fig.~\ref{fighump}. The model has two crystalline ground states, one at high pressure with a hexagonal lattice and a lattice constant of the order
of $r_{\rm min}$. At low pressure the ground state is a more open structure in which the distance $r_{\star}$ appears periodically. At intermediate pressures the system fails to crystallize and forms a glass upon cooling \cite{09BLPZ}.
\begin{figure}
\begin{center}
\includegraphics[width=0.95\linewidth]{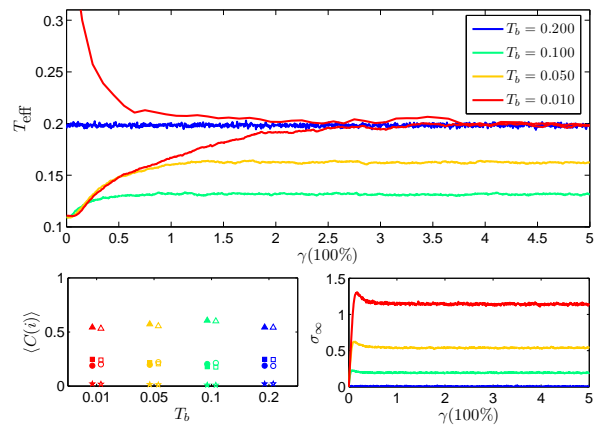}
\caption{Upper panel: trajectories of $T_{\rm eff}$ for the hump model as a function time for a fixed strain rate
$\dot\gamma=10^{-4}$ and $N=6400$, settling at the steady state either from above
or from below. Right lower panels; representative stress-strain curves for the seme model for different bath temperatures.  Lower left panel: The measured steady-state concentration of the four relevant quasi-species of the hump model (full symbols), compared with predictions of Eq. (\ref{CiT}) but with $T_{\rm eff}$ (empty symbols). Circles, triangles, squares and pentagons correspond to $\langle C_i\rangle$ with $i=3,4,5$ and 6 respectively. }
\end{center}
\label{results}
\end{figure}

{\bf Simulations of the elasto-plastic regime:} The molecular dynamics protocol we implemented for both models is
the same.  First, we carefully equilibrated a large number of independent configurations with $N$ particles ($N$ varied
in these simulations between 1024 and 6400) in the NVT ensemble using the Berendsen thermostat over a wide range of temperatures.  These samples were then used to determine
the enthalpies and degeneracies for the two models as described in detail in \cite{09BLPZ}.  These measured
parameters predict accurately the concentrations of quasi species at any given temperature as well as the $\tau_\alpha$ relaxation time.  After this, we turned our equilibrium super--cooled liquids into amorphous solids by minimizing their potential energy (conjugate gradient algorithm), allowing us to sample a representative set of meta-stable minima.  This procedure can be thought of as quenching a liquid infinitely fast into a disordered solid whose temperature is formally $T=0$.  At this point we bring the particle velocities, using a short NVT run, to a value consistent with a desired bath temperature $T_b$. Then we force the system at a constant strain rate $\dot\gamma$ using the SLLOD algorithm combined with Lees--Edwards boundary conditions.  The stress-strain curves obtained for different bath temperatures are shown in the upper panel of Fig. \ref{results} for the hump model.
\begin{figure}
\begin{center}
\includegraphics[width=0.8\linewidth]{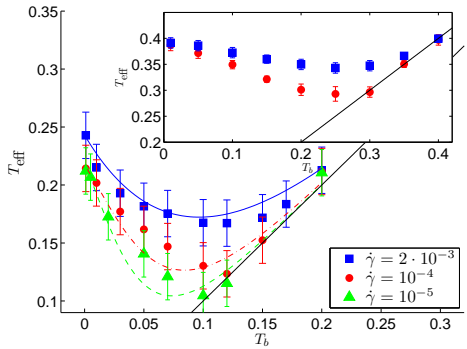}
\caption{Symbols: $T_{\rm eff}$ a s function of $T_b$ for the hump model at three values of the strain rate. Lines: prediction of $T_{\rm eff}$ using Eq. (\ref{scfun}). Inset: $T_{\rm eff}$ a s function of $T_b$ for the Shintani-Tanaka model at the two higher values of the strain rate. Straight lines indicate where
$T_{\rm eff}=T_b$. }
\label{figTeff}
\end{center}
\end{figure}
After the usual elastic response, an irreversible plastic flow begins, eventually generating a time independent steady plastic flow in which all the thermodynamic quantities (such as flow-stress or energy) reached a value independent of the initial quenched configurations.  The major finding is that throughout the evolution the concentrations of quasi-species obey Eq. (\ref{CiT}) with the equilibrium measured values of $g_i$ and ${\cal H}_i$, but with $T_b$ replaced by $T_{\rm eff}$ which is a function of $T_b$ and $\dot \gamma$.  It is remarkable that there is an equally well defined effective temperature not only in the steady state but also in the transient regime. The steady-state values of $T_{\rm eff}$ as a function of $T$ are shown for various values of $\dot \gamma$ in Fig. \ref{figTeff}. Note that at high temperatures
$T_{\rm eff}\to T_b$ whereas $T_{\rm eff}$ increases when $T_b\to 0$, increasing the fluidity of the system.
\begin{figure}
\begin{center}
\includegraphics[width=0.7\linewidth]{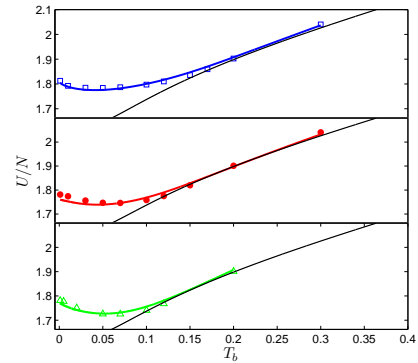}
\caption{Energy per particle in the elasto-plastic steady state as a function of the bath temperature.
Symbols: simulation results. Lines through the symbols: theoretical prediction using Eq. (\ref{U/N}). Black lines:
equilibrium energy per particle. }
\label{UvsTeff}
\end{center}
\end{figure}

The relevance and importance of the steady-state effective temperature can be demonstrated by relating it to the mean energy and the flow stress, the latter being the mean stress in the elasto-plastic steady state. The mean energy per particle $U/N$ was directly measured in the steady state of the hump model, and compared to the theoretical prediction
\begin{equation}
\frac{U}{N} = \frac{0.835}{2} \sum_i \langle C_i\rangle (T_{\rm eff}) i +k_BT_b +\frac{\sigma^2_\infty (T_b)V}{2\mu N} \ ,
\label{U/N}
\end{equation}
where the first two terms were taken from the equilibrium theory for the hump model in Ref. \cite{09BLPZ} but with
$T_{\rm eff}$ replacing $T_b$ in determining the concentrations of the quasi-species; the last
terms is the elastic energy per particles stored in the steady state. The almost perfect agreement between the theoretical
expectation based on $T_{\rm eff}$ and the direct measurement is demonstrated in Fig. \ref{UvsTeff}.
\begin{figure}
\begin{center}
\includegraphics[width=0.8\linewidth]{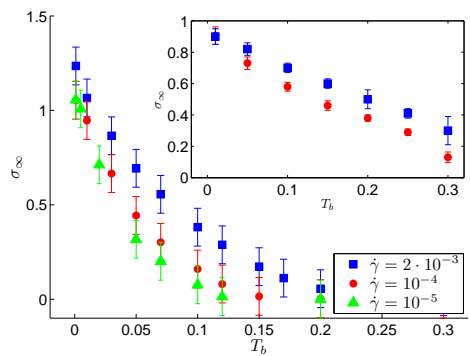}
\caption{Flow stress as a function of $T_b$ for the indicated values of the strain rate $\dot\gamma$ for the hump model, and in the inset for the Shintani-Tanaka model for the two higher values of the strain-rate. }
\label{sigmainf}
\end{center}
\end{figure}

\begin{figure}
\begin{center}
\includegraphics[width=0.8\linewidth]{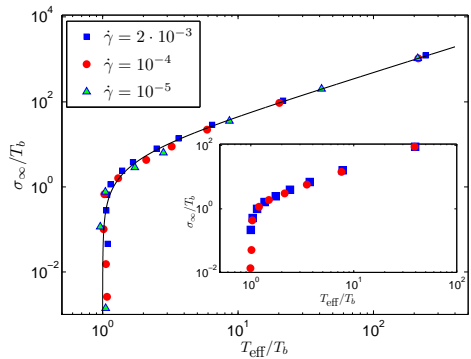}
\caption{Double logarithmic test of the scaling function Eq. (\ref{scfun}); all the data from Figs. \ref{figTeff} and
\ref{sigmainf} are re-plotted here for both models (with the Shintani-Tanaka model in the inset), to demonstrate the excellent data collapse. The continuous line is the function $f(x) = 4.93(x-1)$}
\label{scale}
\end{center}
\end{figure}

The dependence of the flow-stress on $T_b$ for
various values of $\dot\gamma$ is shown in Fig. \ref{sigmainf}.
We show now that we can predict the values of the flow stress $\sigma_\infty(T_b,\dot\gamma)$ given the data for $T_{\rm eff}(T_b,\dot\gamma)$ or vice versa, predict $T_{\rm eff}(T_b,\dot\gamma)$ from the knowledge of $\sigma_\infty(T_b,\dot\gamma)$. To this aim we invoke scaling concepts, and propose a scaling form for $\sigma_\infty(T_b,T_{\rm eff})$ such that the $\dot\gamma$ dependence is carried here by $T_{\rm eff}$:
\begin{equation}
\sigma_\infty(T_b,T_{\rm eff}) = \frac{\rho}{m} k_B T_b ~f\left(\frac{T_{\rm eff}}{T_b}\right) \ , \label{scfun}
\end{equation}
where $\rho$, $m$ are the density and molecular weight of the particles. The function $f(x)$ is a dimensionless scaling function that must obey $f(1)=0$ to agree with the observed fact that at higher temperatures where $T_{\rm eff}=T_b$ the flow stress approaches zero. For $T\to 0$ or $x\to \infty$ we observed that $\sigma_\infty$ becomes proportional to $T_{\rm eff}$ requiring $f(x)\to C x$ for $x\to \infty$.
The simplest function that obeys these limits is $f(x) = C(x-1)$. A test of the predicted data collapse is
shown in Fig. \ref{scale} where the continuous line is the function $4.93(x-1)$.
Having the scaling function at hand we can predict the data, say of $T_{\rm eff} (T_b,\dot\gamma)$ from the knowledge
of $\sigma_\infty(T_b,\dot\gamma)$, or vice versa.  Using for example the data for $\sigma_\infty$ for the hump model in Fig. \ref{sigmainf} and the scaling function
$f(x) =4.93(x-1)$ we solve for $T_{\rm eff} (T_b,\dot\gamma)$. The results are demonstrated in Fig. \ref{figTeff} with the curved lines going through the data. We conclude that the procedure is in satisfactory agreement with the data, demonstrating the importance of the concept of effective temperature. We mention in passing that the temperature
$T^*$ where $T_{\rm eff}$ separates from $T_b$ can be easily predicted by equating the relaxation rate $\tau_\alpha(\xi(T))$ with $\dot\gamma^{-1}$: $T^*\sim \ln^{-1}{\tau_0\dot\gamma}$. At temperatures higher than $T^*$ the natural relaxation time $\tau_\alpha$ is the shorter of the two, whereas at temperatures lower than the minimum the shear rate determines the rate of relaxation.

Much remains to be done. For example in the case of the standard model of
binary mixtures we did not find a satisfactory up-scaling that
remains the same in and out of equilibrium. This and other riddles concerning the present
approach will be dealt with in a future publication. Nevertheless we believe that the present findings will provide
grounds for new exciting future research.

This work has been supported in part by the Israel Science Foundation and the German Israeli Foundation.


\begin{thebibliography}{99}

\bibitem{79AK}
A.S. Argon and H.Y. Kuo, Mater. Sci. Eng. {\bf 39}, 101 (1979).

\bibitem{79Arg}
A.S. Argon, Acta Metall. {\bf 27}, 47 (1979).

\bibitem{82AS}
A.S. Argon and L. T. Shi, Philos. Mag. A {\bf 46}, 275 (1982).

\bibitem{98FL}
M.L. Falk and J.S. Langer, Phys. Rev. E {\bf 57}, 7192 (1998).

\bibitem{98Sol}
P. Sollich, Phys. Rev. E, {\bf 58}, 738 (1998).

\bibitem{07BLP}
E. Bouchbinder, J.S. Langer and I. Procaccia, Phys. Rev. E, {\bf 75}, 036107 (2007); {\bf 75}, 036108 (2007).

\bibitem{09BCA}
L. Bocquet, A. Colin and A. Ajdari, Phys. Rev. Lett. {\bf 103}, 036001 (2009).

\bibitem{97SLHC}
P. Sollich, F. Lequeuz, P. H\'ebrand and M. E. Cates, Phys. Rev. Lett. {\bf 78}, 2020 (1997).

\bibitem{09BL}
E. Bouchbinder and J.S. Langer, arXiv:0903.1524v1 [cond-mat.mtr-sci] 9 May (2009).



\bibitem{89Ed}
S. F. Edwards and R. B. S. Oakeshott Physica A {\bf 157}, 1080-1090 (1989).

\bibitem{98Cug}
L. Cugliandolo in H Falomir et al, Am. Inst. Phys. Conference Proceedings of the 1998 Buenos Aires meeting.


\bibitem{07BK}
L. Bou\'e and E. Katzav Euro. Phys. Lett. {\bf 80} 54002 (2007);
S. Deboeuf, M. Adda--Bedia and A. Boudaoud Euro. Phys. Lett. {\bf 85} 24002 (2009).



\bibitem{02BB}
L. Berthier and J.-L. Barrat, Phys. Rev. Lett. {\bf 89}, 095702 (2002).

\bibitem{04OLN}
C. S. O'Hern, A. J. Liu, and S. R. Nagel, Phys. Rev. Lett. 93, 165702 (2004).

\bibitem{07HL}
T. Haxton and A. J. Liu, Phys. Rev. Lett. 99, 195701 (2007).

\bibitem{Chetrite}
J. R. Gomez-Solano, A.Petrosyan, S. Ciliberto, R. Chetrite and K. Gawedzki, Phys. Rev. Lett.  {\bf 103}, 040601 (2009).

\bibitem{series}
E. Aharonov, E. Bouchbinder, V. Ilyin, N. Makedonska, I. Procaccia and N. Schupper,
Europhys. Lett. {\bf 77}, 56002 (2007); H. G. E. Hentschel, V. Ilyin, N. Makedonska, I. Procaccia and N. Schupper, Phys. Rev. E {\bf 75}, 050404 (2007); E. Lerner and I. Procaccia, Phys. Rev. E {\bf 78}, 020501 (2008); E. Lerner, I. Procaccia, and J. Zylberg Phys. Rev. Lett. {\bf 102}, 125701 (2009); H. G. E. Hentschel, V. Ilyin and I. Procaccia, Phys. Rev. Lett. {\bf 101} 265701 (2008); H. G. E. Hentschel, V. Ilyin, I. Procaccia and N. Schupper,  Phys. Rev. E, {\bf 78} 061504 (2008).

\bibitem{09BLPZ}
L. Bou\'e, E. Lerner, I. Procaccia, J. Zylberg, ``Predictive Statistical Mechanics for Glass Forming Systems", J. Stat. Mech, submitted. ArXiv: arXiv:0905.3962

\bibitem{07EP}
J.P. Eckmann and I. Procaccia, Phys. Rev. E,

\bibitem{06ST}
H. Shintani and H. Tanaka, Nature Physics {\bf 2}, 200 (2006).

\bibitem{89Dzu}
M. Dzugutov, Phys. Rev. A 40, 5434 (1989).


\bibitem{07ILLP}
V. Ilyin, E. Lerner, T-S Lo and I. Procaccia, Phys. Rev.
Lett., {\bf 99}, 135702 (2007).

\bibitem{09LPR}
 E. Lerner, I. Procaccia and I. Regev, Phys. Rev E, {\bf 79}, 031501 (2009).


\end{thebibliography}
\end{document}